\documentclass[aps,preprint,nofootinbib]{revtex4}%
\usepackage{amsfonts}
\usepackage{amsmath}
\usepackage{amssymb}
\usepackage{graphicx}%
\providecommand{\U}[1]{\protect\rule{.1in}{.1in}}

\begin{document}
\preprint{ }
\title[Short title for running header]{Remarks on the \textquotedblleft non-canonicity puzzle\textquotedblright:
Lagrangian symmetries of the Einstein-Hilbert action}
\author{N. Kiriushcheva}
\email{nkiriush@uwo.ca}
\author{P.G. Komorowski}
\email{pkomoro@uwo.ca}
\author{S.V. Kuzmin}
\email{skuzmin@uwo.ca}
\affiliation{The Department of Applied Mathematics, The University of Western Ontario,
London, Ontario, Canada, N6A 5B7}
\pacs{PACS number}

\begin{abstract}
Given the non-canonical relationship between variables used in the Hamiltonian
formulations of the Einstein-Hilbert action (due to Pirani, Schild, Skinner
(PSS) and Dirac) and the Arnowitt-Deser-Misner (ADM) action, and the
consequent difference in the gauge transformations generated by the
first-class constraints of these two formulations, the assumption that the
Lagrangians from which they were derived are equivalent leads to an apparent
contradiction that has been called \textquotedblleft the non-canonicity
puzzle\textquotedblright. In this work we shall investigate the group
properties of two symmetries derived for the Einstein-Hilbert action:
diffeomorphism, which follows from the PSS and Dirac formulations, and the one
that arises from the ADM formulation. We demonstrate that unlike the
diffeomorphism transformations, the ADM transformations (as well as others,
which can be constructed for the Einstein-Hilbert Lagrangian using Noether's
identities) do not form a group. This makes diffeomorphism transformations
unique (the term \textquotedblleft canonical\textquotedblright%
\ symmetry\ might be suggested). If the two Lagrangians are to be called
equivalent, canonical symmetry must be preserved. The interplay between
general covariance and the canonicity of the variables used is discussed.

\end{abstract}
\volumeyear{year}
\volumenumber{number}
\issuenumber{number}
\eid{identifier}
\date{\today}
\received{}

\maketitle


\section{Introduction}

An analysis of the two oldest Hamiltonian formulations of the second-order
Einstein-Hilbert (EH) action for metric General Relativity (i.e. Pirani,
Schild, and Skinner (PSS) \cite{PSS}; and Dirac \cite{Dirac}), was completed
in \cite{KKRV, Myths}. Using the approach of Castellani \cite{Castellani}, it
was demonstrated that first-class constraints produce a\ generator for the
diffeomorphism invariance\footnote{We understand diffeomorphism invariance
(\textit{diff}) as \textquotedblleft active\textquotedblright\ \cite{Rovelli}
(p. 62) when \textquotedblleft coordinates play no role\textquotedblright,
i.e. transformations of fields written in the same coordinate system.} - the
known gauge symmetry of the Einstein-Hilbert (EH) action. This outcome
contradicts the result of the Arnowitt, Deser, and Misner formulation (ADM, or
geometrodynamics) \cite{ADM, goldies} where the constraints lead to a
different symmetry, one which is known by many names: \textquotedblleft
spatial diffeomorphism\textquotedblright,\ \textquotedblleft special induced
diffeomorphism\textquotedblright,\ \textquotedblleft field-dependent
diffeomorphism\textquotedblright,\ \textquotedblleft foliation preserving
diffeomorphism\textquotedblright, \textquotedblleft one-to-one
correspondence\textquotedblright, \textquotedblleft one-to-one
mapping\textquotedblright\ (see \cite{Myths} and references therein). It was
shown \cite{FKK, Myths} that the PSS and Dirac Hamiltonians are related by a
canonical transformation of the phase-space variables; while the
transformation from the Dirac to ADM Hamiltonian is not a canonical change of
variables. Canonical transformations must preserve all properties of the
Hamiltonian. Because gauge symmetry is an important characteristic of a
constrained system, a difference between the symmetries of the PSS (or Dirac)
and ADM Hamiltonian formulations indicates that a non-canonical relationship
exists between the two formulations. This truism was explicitly confirmed in
\cite{FKK, Myths} by the calculation of Poisson brackets (PBs) among the
phase-space variables. Waxing poetic, the term \textquotedblleft the
non-canonicity puzzle\textquotedblright\ was coined in \cite{CLM} to describe
the results of \cite{FKK, Myths}; but in \cite{ShestakovaCQG}, using more
direct language, it is called \textquotedblleft the contradiction that again
witnesses about the incompleteness of the theoretical
foundations\textquotedblright. The source of the \textquotedblleft
puzzle\textquotedblright\ or \textquotedblleft contradiction\textquotedblright%
\ lies in finding how to reconcile the non-equivalence of the two Hamiltonian
formulations with their corresponding Lagrangian formulations when
\textquotedblleft it is supposed\textquotedblright\ \cite{ShestakovaGandC} or
\textquotedblleft it is believed\textquotedblright(as in \cite{ShestakovaCQG})
\textquotedblleft that each of them is equivalent to the Einstein (Lagrangian)
formulation\textquotedblright\ \cite{ShestakovaGandC, ShestakovaCQG}.

This belief might lead one to conclude that Dirac's Hamiltonian formulation of
constrained systems is incomplete \cite{ShestakovaCQG}, and proposals ought to
follow on how to redefine the primary constraints, on how to use boundary
terms, on how to have \textquotedblleft two non-canonical transformations that
compensate each other\textquotedblright\ \cite{CLM}, and on how to modify the
PBs through the extension of phase space\footnote{Such an extension is
obtained by replacing the classical EH action by \textquotedblleft the
effective action including gauge and ghost sectors\textquotedblright%
\ \cite{ShestakovaGandC, ShestakovaCQG}.} \cite{ShestakovaGandC,
ShestakovaCQG}. But such approaches immediately give rise to a general
question: why are such manipulations needed for one formulation (i.e. ADM),
but not for the others (i.e. PSS and Dirac)? Hamiltonian solutions of the
\textquotedblleft puzzle\textquotedblright\ proposed in \cite{CLM,
ShestakovaGandC, ShestakovaCQG}\ deserve a more detailed discussion; but in
this article we shall limit ourselves to the consideration of the Lagrangian
formulation and the symmetries of the EH\footnote{In PSS \cite{PSS} the
gamma-gamma part of the EH action is considered; and Dirac in \cite{Dirac}
made some additional manipulations in this Lagrangian. In spite of these
modifications, both formulations lead to exactly the same equations of motion
as the original EH action; and the metric is an independent field-variable in
all of them.} and ADM Lagrangians. We note that the literature on the
Hamiltonian formulation of constrained systems contains various treatments
with claims that the variables, which appear in the original Lagrangian, have
different properties (i.e. they might be dynamical and non-dynamical; or some
variables can be treated as Lagrange multipliers; or some variables are
canonical, but some not; or some have conjugate momenta, but others do not
need them; or that secondary first-class constraints can be \textquotedblleft
promoted\textquotedblright\ to primary first-class constraints, et cetera (see
\cite{Myths} and references therein)). This plethora of treatments through
which different results may be obtained, depending on the ingenuity of an
investigator, leave an impression that the Hamiltonian method for constrained
systems\textit{ }is an art, not a defined and unambiguous procedure (or, as
suggested in \cite{CLM, ShestakovaGandC, ShestakovaCQG} that it is not yet a procedure).\ 

Instead of relying upon the first-class constraints, as in the Dirac
procedure, one may use the Lagrangian formulation of a singular system to
derive symmetries from the Noether identities. The differential relationships
among the Euler-Lagrange derivatives are linked to the gauge transformations;
thus the treatment of the Lagrangian is free from the artistic approaches that
have been applied to the Hamiltonian. Lagrangian symmetries describe a
transformation in which \textit{all} fields are treated on the same footing,
irrespective of the name assigned to them (e.g. \textquotedblleft
dynamical\textquotedblright\ and \textquotedblleft
non-dynamical\textquotedblright, et cetera). The differential identities (DIs)
involve\textit{ }the Euler-Lagrange derivatives with respect to \textit{all}
fields. We must emphasize that the Lagrangian and Hamiltonian methods should
have the same mathematical rigor; but the main reason for us to consider
Lagrangian symmetries is to aid in developing a criterion for the equivalence
of two Lagrangians in light of this \textquotedblleft puzzle\textquotedblright.

In \cite{Myths}, an analysis of the Hamiltonian formulations of Dirac and ADM
was performed; it was concluded that if two Hamiltonian formulations are not
related by a canonical transformation and if they have different symmetries
(i.e. they are not equivalent), then the corresponding Lagrangians are also
not equivalent, contrary to the \textquotedblleft belief\textquotedblright%
\ which forms the basis of the \textquotedblleft puzzle\textquotedblright. If
two Lagrangians are not equivalent, then the results of \cite{FKK, Myths} are
fully consistent, there is no puzzle, and the theoretical foundations are
sound. The conclusion drawn that PSS and Dirac are not equivalent to the ADM
formulation was based on the belief of the authors of \cite{Myths} that
Dirac's Hamiltonian method for constrained systems is an unambiguous
procedure, applicable to any theory, that leads to a unique symmetry that
corresponds exactly to the symmetry present in the Lagrangian.

The change of variables used by ADM \cite{ADM, goldies} to go from the metric
tensor $g_{\mu\nu}$ of the EH action (used in PSS and Dirac) to the ADM
variables (lapse $N$, shift $N^{i}$ and space-space components of the metric
tensor $\gamma_{km}$)\footnote{We employ Greek letters for space-time indices:
$\mu=0,1,2,3$. Latin letters for space indices: $k=1,2,3$, and
\textquotedblleft$0$\textquotedblright\ for the time index.} is%

\begin{equation}
N=\left(  -g^{00}\right)  ^{-1/2}\text{ },\text{\ }N^{i}=-\frac{g^{0i}}%
{g^{00}}\text{ \ },\gamma_{km}=g_{km}~,\label{eqnL1}%
\end{equation}
which is invertible, but not covariant. It is this condition of invariability
that some view as sufficient for the EH and ADM Lagrangians to be equivalent.
But this conclusion is not an obvious one to make when dealing with singular
Lagrangians. In the Hamiltonian formulation of a singular and covariant
Lagrangian, gauge symmetries are derived from the first-class constraints; at
the Lagrangian level, gauge symmetries are associated with the existence of
the Noether identities \cite{Noether}\footnote{For an English translation of
Noether's paper see \cite{Noether-eng}.}. The invariability of redefinition
(\ref{eqnL1}) allows one, starting from known transformations of one set of
variables (e.g. $\delta_{\mathit{diff}}\left\{  g_{\mu\nu}\right\}  $), to
find the transformations for another set (i.e. $\delta_{\mathit{diff}}\left\{
N,N^{i},\gamma_{km}\right\}  $), and \textit{vise} \textit{versa} (i.e. from
$\delta_{\mathit{ADM}}\left\{  N,N^{i},\gamma_{km}\right\}  $ one finds
$\delta_{\mathit{ADM}}\left\{  g_{\mu\nu}\right\}  $). For example, one can
also check whether $\delta_{\mathit{ADM}}$, which is a symmetry of the ADM
Lagrangian, is also a symmetry of the EH Lagrangian (i.e. whether the EH
Lagrangian is invariant under transformations given by $\delta_{\mathit{ADM}%
}\left\{  g_{\mu\nu}\right\}  $). Direct calculation is difficult; but by the
converse of Noether's theorem \cite{Noether} (i.e. if an action is invariant
under some symmetry, then there exists the corresponding DI) and by assuming
that $\delta_{\mathit{ADM}}\left\{  g_{\mu\nu}\right\}  $ is an invariance,
one can find the corresponding DI\footnote{Such constructions were described
by Schwinger \cite{Schwinger}, see also \cite{Trans}.} and directly check it.
A shorter approach is to connect a new DI to a known DI of the EH action; in
addition, any DIs that are independent linear combinations of known DIs also
describe symmetries. In such a way it is not difficult to demonstrate that
$\delta_{\mathit{ADM}}\left\{  g_{\mu\nu}\right\}  $ is indeed a symmetry of
the EH action. Conversely, one starting from $\delta_{\mathit{diff}}\left\{
g_{\mu\nu}\right\}  $ can find $\delta_{\mathit{diff}}\left\{  N,N^{i}%
,\gamma_{km}\right\}  $ and check that it is a symmetry of the ADM Lagrangian
by\ using its DI (e.g. see p. 17 of \cite{BGMR}). In a similar way, linear
combinations of the DIs can be used to construct other transformations that
will be symmetries of both the EH and ADM Lagrangians. But do these
relationships prove the equivalence of the EH and ADM Lagrangians? One may
also ask why it is that when using the Hamiltonian method, one particular
symmetry follows from the constraint structure, but in the Lagrangian, we
apparently have an infinity of equally good symmetries? Such a non-uniqueness
should be a warning sign\footnote{Of course, these questions can be avoided
assuming that \textquotedblleft Hamiltonian dynamics is not completely
equivalent to Lagrangian formulation of the original theory. In Hamiltonian
formalism the constraints generate transformations of phase-space variables;
however, the group of these transformations does not have to be equivalent to
the group of gauge transformations of Lagrangian theory\textquotedblright%
\ \cite{Shestakova}. Such an assumption eliminates a \textquotedblleft
non-canonicity puzzle\textquotedblright\ or, alternatively, it provides the
solution: the EH and ADM Lagrangians are equivalent, but the corresponding
Hamiltonians just happen to pick different symmetries.}.

The key concept that allows one to distinguish among the numerous symmetries
due to the Lagrangian approach may be found in \cite{ShestakovaCQG}, where it
is suggested that Dirac's method is incomplete. According to
\cite{ShestakovaCQG}, \textquotedblleft the difference in the \textit{groups}
of transformations is the first indication to the inconsistency of the
theory\textquotedblright\textit{(italic is ours)}. This key concept can be
used to answer the question about how to classify all symmetries that can be
constructed for \textit{one} Lagrangian: which of the symmetries have group
properties and thus constitute the \textquotedblleft basic\textquotedblright,
\textquotedblleft true\textquotedblright,\ or \textquotedblleft
canonical\textquotedblright\ symmetries? This also allows one to compare
\textit{two} Lagrangian formulations by matching those symmetries that have
group properties for each. A related question is: if only one symmetry has the
property to form a group, is it the symmetry that the Hamiltonian formulation
produces (or should produce)? At least for the EH Lagrangian, where
diffeomorphism is a gauge symmetry with a group property \cite{Bergmann}, its
Hamiltonian formulation \cite{PSS, Dirac} leads exactly to this symmetry
\cite{KKRV, Myths} without any extension of Dirac's procedure. Now consider
going from one Lagrangian to another by performing some invertible change of
variables. If the symmetry that had a group property in the original
formulation ceases to have a group property in a new formulation, but another
symmetry that did not have group properties in the original formulation
\textquotedblleft develops\textquotedblright\ group properties in a new
formulation, is this \textquotedblleft the first indication to the
inconsistency of the theory\textquotedblright\ \cite{ShestakovaCQG} or is it
proof that the two Lagrangian formulations are not equivalent? Perhaps a
change of variables that creates such a result should be called
\textquotedblleft non-canonical\textquotedblright,\ in analogy with
Hamiltonian terminology where a non-canonical change of variables also causes
a change of symmetries \cite{FKK, Myths}.

The investigation of the symmetries of the two Lagrangians (EH and ADM), which
are related to each other by the change of variables (\ref{eqnL1}), is a less
cumbersome calculation to perform compared with the Hamiltonian method; the
same is true of the study of whether a symmetry with a group property, of one
Lagrangian, is also a symmetry with a group property, of the other Lagrangian.
But for non-covariant changes such an investigation is complicated. In
particular, to identify a symmetry with a group property, the commutators of
two transformations must be considered, and in the case of field-dependent
structure functions, higher, nested commutators are needed. \ For
non-covariant variables these calculations must be performed separately for
different fields, and the consistency of different commutators must be
checked. In this article we discuss the simple parts of the calculation that
one may perform in a quasi-covariant form, and consider two symmetries of the
EH Lagrangian: transformations of the metric tensor $g^{\mu\nu}$ under
\textit{diff} ($\delta_{\mathit{diff}}g^{\mu\nu}$) and under ADM
transformations ($\delta_{\mathit{ADM}}g^{\mu\nu}$). We also compare their
group properties. In the next Section we briefly review some results relevant
to the \textit{diff} invariance of the EH action with an emphasis placed on
the role of DIs, their direct connection to the form of the transformations,
and the possible construction of additional symmetries by using combinations
of DIs (i.e. the results that will be needed for a discussion of
$\delta_{\mathit{ADM}}g^{\mu\nu}$). In Section III we demonstrate the
invariance of the EH action under the ADM transformations $\delta
_{\mathit{ADM}}g^{\mu\nu}$, and show that unlike \textit{diff}, the
$\delta_{\mathit{ADM}}g^{\mu\nu}$ do not constitute a group. More cumbersome
calculations of the group properties of the same transformations of the ADM
variables, $\delta_{\mathit{diff}}\left\{  N,N^{i},\gamma_{km}\right\}  $ and
$\delta_{\mathit{ADM}}\left\{  N,N^{i},\gamma_{km}\right\}  $, for the ADM
Lagrangian are in progress and will be reported elsewhere. In the Conclusion,
we summarize our results and discuss the consequences of\textit{ }%
$\delta_{\mathit{diff}}\left\{  N,N^{i},\gamma_{km}\right\}  $\textit{
}and\textit{ }$\delta_{\mathit{ADM}}\left\{  N,N^{i},\gamma_{km}\right\}  $
either having or not having group properties, in all possible combinations.
Finally we comment on the role of covariance.

\section{Symmetries in the Lagrangian approach of the Einstein-Hilbert action}

There are statements in the literature such as: \textquotedblleft...one of the
advantages of the Hamiltonian formulation is that one does not have to specify
the gauge symmetries \textit{a priori}. Instead, the structure of the
Hamiltonian constraints provides an essentially algorithmic way in which the
correct gauge symmetry structure is determined automatically\textquotedblright%
\ \cite{Horava}. We note that this is not a special or exclusive property of
the Hamiltonian method\footnote{It has to be admitted that applications of
Hamiltonian methods can lead to very long and cumbersome calculations, which
in some cases, are not straightforward.}. The Lagrangian approach also
provides an algorithm, which is due to Noether's second theorem for finding
gauge symmetries \cite{Noether, Noether-eng}, that connects these symmetries
with the DIs - combinations of Euler-Lagrange derivatives that are identically
equal to zero (off-shell). The Hamiltonian method provides an algorithm for
finding and classifying constraints (all first-class constraints are needed to
find a symmetry); in the Lagrangian approach, DIs can be built using an
iterative procedure. For the Einstein-Cartan (EC) action, which has richer
symmetry properties than EH, such a construction was performed in
\cite{Trans}. In the same way, DIs can be built for the EH action
\cite{Myths}. The relative simplicity of such calculations is due to the
covariance of the theories considered. It would be a much more complicated
procedure to try to find DIs in non-covariant theories, or non-covariant DIs
for covariant theories. Of course, for any theory for which there is no
\textit{a priori} knowledge of the existence of gauge symmetries, it is
unproductive to search for identities without preliminary analysis. The first
step is to determine if a Lagrangian is singular, by evaluating its Hessian:%

\begin{equation}
H^{\alpha\beta}=\frac{\delta^{2}L}{\delta\dot{Q}_{\alpha}~\delta\dot{Q}%
_{\beta}}~,\label{eqnL3}%
\end{equation}
where $\dot{Q}_{\alpha}$ are the time derivatives of $Q_{\alpha}$ - the
independent fields of the Lagrangian. If the determinant of the Hessian is
zero, then the Lagrangian is singular; the rank of the Hessian is related to
the number of independent DIs that can be found. It should be noted that
singularity of the Lagrangian is a necessary condition to have a gauge
symmetry, but not sufficient (the simplest example is the massive vector
(Proca) field where the Lagrangian is singular, but has no gauge symmetry).
The rank of the Hessian provides only an upper bound on the maximum number of
independent gauge symmetries. The Hessian is often written for velocities,
even for the Lagrangian of covariant theories; but time is not special for
covariant theories and singling it out is unnecessary.

In the Hamiltonian approach, knowledge of the first-class constraints is
sufficient to restore gauge invariance: for example, by using the Castellani
procedure \cite{Castellani}\textbf{.} Although there are some modifications of
the Castellani procedure, they must be used with care (see
\cite{affine-metric}). Similarly for Lagrangians, if the DIs are known, then
transformations can be found using the explicit connections of the DIs and the
transformations \cite{Noether, Schwinger}. The approach described for finding
\textit{a priori} unknown Lagrangian symmetries is general; but for the EH
action, a well-known covariant DI had already appeared, along with the EH
action\footnote{So, it is not easy to recognize this due to Hilbert's
presentation and also complications with coupling to Mie's electrodynamics. In
addition, this identity was known before any connection was made to
Euler-Lagrange derivatives of the EH action - this is simply the contracted
Bianchi identity \cite{Bianchi}.} itself, in Hilbert's paper \cite{Hilbert}%
\footnote{For an English translation see \cite{Hilbert-eng}.}.

We shall briefly illustrate the application of this general procedure to the
EH action. These results will be needed, used, and compared in the next
Section, where the ADM symmetry is discussed. The Einstein-Hilbert action is
\cite{Landau, Carmeli}%

\begin{equation}
S_{EH}=\int L~d^{4}x=\int\sqrt{-g}R~d^{4}x~, \label{eqnL5}%
\end{equation}
where $g=\det g_{\mu\nu}$, $L$ is the scalar density (Lagrangian density) and
the Ricci scalar $R$, Ricci tensor $R_{\mu\nu}$, and Christoffel symbol
$\Gamma_{\mu\nu}^{\alpha}$ are:%

\begin{equation}
R=g^{\mu\nu}R_{\mu\nu},\text{ \ \ \ \ }R_{\mu\nu}=\Gamma_{\mu\nu,\alpha
}^{\alpha}-\Gamma_{\mu\alpha,\nu}^{\alpha}+\Gamma_{\mu\nu}^{\alpha}%
\Gamma_{\alpha\beta}^{\beta}-\Gamma_{\mu\beta}^{\alpha}\Gamma_{\alpha\nu
}^{\beta}~, \label{eqnL6}%
\end{equation}%
\begin{equation}
\Gamma_{\mu\nu}^{\alpha}=\frac{1}{2}g^{\alpha\beta}\left(  g_{\mu\beta,\nu
}+g_{\nu\beta,\mu}-g_{\mu\nu,\beta}\right)  . \label{eqnL7}%
\end{equation}
The variational, Euler-Lagrange derivative (ELD) of the EH action is%

\begin{equation}
E^{\alpha\beta}=\frac{\delta L_{EH}}{\delta g_{\alpha\beta}}=\sqrt{-g}\left(
\frac{1}{2}g^{\alpha\beta}R-R^{\alpha\beta}\right)  =-\sqrt{-g}G^{\alpha\beta
}, \label{eqnL9}%
\end{equation}
where $G^{\alpha\beta}=R^{\alpha\beta}-\frac{1}{2}g^{\alpha\beta}R$ is the
Einstein tensor.

It is not difficult to find the DI by using a general construction similar to
that performed for the Einstein-Cartan action \cite{Trans}. Under the
reasonable assumption that a covariant theory should also have covariant
identities, and given the rank of the Hessian for the EH action, the DI
follows almost immediately. The rank of the Hessian is six; and because the
second-rank metric tensor has ten independent components, there should be four
independent DIs. The four covariant identities that one can build from the
ELDs, which are covariant symmetric second-rank tensor densities, consist of
either four scalars or one vector. It is impossible to construct four scalars
from the ELDs; but to find a true vector, one may take a covariant derivative
of a second-rank tensor to yield:%

\begin{equation}
I^{\mu}=E_{;\nu}^{\mu\nu}=E_{,\nu}^{\mu\nu}+\Gamma_{\alpha\beta}^{\mu
}E^{\alpha\beta}\equiv0.\label{eqnL15}%
\end{equation}
By direct substitution, one can easily confirm that this combination is
identically zero.

Schwinger's paper \cite{Schwinger} contains a description of how to construct
the DIs from known gauge transformations; and in correspondence to Noether's
theorem, this process also applies in inverse order, through\ the converse
relationship between DIs and transformations. One forms a scalar from a vector
DI (\ref{eqnL15}) by using the gauge parameters of appropriate tensorial
dimension followed by equating the scalar to variations of the action. \ One writes%

\begin{equation}
\delta S_{EH}=\int\delta g_{\mu\nu}E^{\mu\nu}d^{4}x=\int\xi_{\mu}I^{\mu}%
d^{4}x~, \label{eqnL17}%
\end{equation}
and performing integration by parts yields%
\begin{equation}
\int\xi_{\mu}I^{\mu}d^{4}x=\int\xi_{\mu}\left(  E_{,\nu}^{\mu\nu}%
+\Gamma_{\alpha\beta}^{\mu}E^{\alpha\beta}\right)  d^{4}x=\int\left(
-\frac{1}{2}\xi_{\mu,\nu}-\frac{1}{2}\xi_{\nu,\mu}+\Gamma_{\mu\nu}^{\alpha}%
\xi_{\alpha}\right)  E^{\mu\nu}d^{4}x~, \label{eqnL18}%
\end{equation}
then one obtains%

\begin{equation}
\delta_{\mathit{diff}}g_{\mu\nu}=-\frac{1}{2}\xi_{\mu,\nu}-\frac{1}{2}\xi
_{\nu,\mu}+\Gamma_{\mu\nu}^{\alpha}\xi_{\alpha}~.\label{eqnL19}%
\end{equation}
Note that the coefficient $\frac{1}{2}$ also appears when symmetries are
restored in the Hamiltonian approach \cite{KKRV, Myths} but this result is
usually presented in a different form. The constant $\frac{1}{2}$ can be
incorporated into a gauge parameter without any effect on the results; and we
will use the shorter form%

\begin{equation}
\delta_{\mathit{diff}}g_{\mu\nu}=-\xi_{\mu,\nu}-\xi_{\nu,\mu}+2\Gamma_{\mu\nu
}^{\alpha}\xi_{\alpha}=-\xi_{\mu;\nu}-\xi_{\nu;\mu}~, \label{eqnL20}%
\end{equation}
which is a manifestly covariant expression, a consequence of using a covariant
DI. Knowledge of this transformation allows one to also find transformations
for any combination built from the metric, for example,%

\begin{equation}
\delta_{\mathit{diff}}\Gamma_{\mu\nu}^{\alpha}=-\xi^{\beta}\Gamma_{\mu
\nu,\beta}^{\alpha}+\Gamma_{\mu\nu}^{\beta}\xi_{,\beta}^{\alpha}-\Gamma
_{\mu\beta}^{\alpha}\xi_{,\nu}^{\beta}-\Gamma_{\nu\beta}^{\alpha}\xi_{,\mu
}^{\beta}-\xi_{,\mu\nu}^{\alpha}~, \label{eqnL25}%
\end{equation}%
\begin{equation}
\delta_{\mathit{diff}}R_{\mu\nu}=-\xi^{\rho}R_{\mu\nu,\rho}-\xi_{,\mu}^{\rho
}R_{\nu\rho}-\xi_{,\nu}^{\rho}R_{\mu\rho}~,\left.  {}\right.  \left.
{}\right.  \delta R=-\xi^{\rho}R_{,\rho}~, \label{eqnL26}%
\end{equation}
and%

\begin{equation}
\delta_{\mathit{diff}}G_{\mu\nu}=-\xi^{\rho}G_{\mu\nu,\rho}-\xi_{,\mu}^{\rho
}G_{\nu\rho}-\xi_{,\nu}^{\rho}G_{\mu\rho}~.\label{eqnL27}%
\end{equation}
We note that gauge parameters in general, and the $\xi_{\mu}$ of the EH action
in particular, are field-independent, as was explicitly stated by Hilbert
\cite{Hilbert}, Noether \cite{Noether}, and others, and by Rosenfeld, in the
first discussion on the Hamiltonian formulation for a singular Lagrangian
\cite{Rosenfeld}\footnote{For an English translation see \cite{Preprint}.}.
Further, the methods used to restore gauge symmetries, such as the Castellani
approach \cite{Castellani}, are also based on the condition that the gauge
parameters be field-independent.

Our goal is to compare the \textit{diff }transformations with the ADM
transformations of the same EH action; therefore, transformations of the same
variables (i.e. the metric tensor) under ADM are needed. For this purpose we
find them from the transformations of ADM variables by using their connection
to the metric (\ref{eqnL1}). The lapse and shift are expressed in terms of
contravariant components, so it is easier to find the transformations of
contravariant components of the metric from the ADM transformations. Of
course, transformations of covariant and contravariant components of the
metric have a simple relationship due to $g_{\mu\nu}g^{\nu\alpha}=\delta_{\mu
}^{\alpha}$; but for our discussion, it is preferable to know the DIs that
lead directly to $\delta_{\mathit{diff}}g^{\mu\nu}$. There are a few ways to
find a DI that is expressed in terms of a covariant ELD; we might use a DI
that is already known, and consider%

\begin{equation}
I_{\alpha}=g_{\alpha\mu}I^{\mu}~, \label{eqnL30}%
\end{equation}
which is also an identity. After performing some simple rearrangement, we obtain%

\begin{equation}
I_{\alpha}=g_{\alpha\mu}I^{\mu}=-2\left(  g^{\mu\nu}E_{\mu\alpha}\right)
_{,\nu}-g_{,\alpha}^{\mu\nu}E_{\mu\nu}~. \label{eqnL31}%
\end{equation}
Repeating steps (\ref{eqnL17}) - (\ref{eqnL20}) for a covariant DI,
$I_{\alpha}$, and using%

\begin{equation}
\delta S_{EH}=\int\delta g^{\mu\nu}E_{\mu\nu}d^{4}x=\int\xi^{\alpha}I_{\alpha
}d^{4}x~, \label{eqnL32}%
\end{equation}
we obtain%

\begin{equation}
\delta_{diff}g^{\mu\nu}=\xi_{,\alpha}^{\nu}g^{\mu\alpha}+\xi_{,\alpha}^{\mu
}g^{\nu\alpha}-g_{,\alpha}^{\mu\nu}\xi^{\alpha} \label{eqnL33}%
\end{equation}
(here we also incorporated the constant $\frac{1}{2}$ into the gauge parameter).

But do these transformations form a group? To answer this question, the
commutator of two transformations is needed (i.e. $\left[  \delta_{2}%
,\delta_{1}\right]  $) for which it should be possible to present the result
in the form of a single transformation, but with a new parameter
$\delta_{\left[  1,2\right]  }$, %

\begin{equation}
\left[  \delta_{2},\delta_{1}\right]  g^{\mu\nu}=\left(  \delta_{2}\delta
_{1}-\delta_{1}\delta_{2}\right)  g^{\mu\nu}=\delta_{\left[  1,2\right]
}g^{\mu\nu}~. \label{eqnL40}%
\end{equation}
This result was found by Bergmann and Komar \cite{Bergmann} in the following form:%

\begin{equation}
\xi_{\left[  1,2\right]  }^{\alpha}=\xi_{2}^{\beta}\xi_{1,\beta}^{\alpha}%
-\xi_{1}^{\beta}\xi_{2,\beta}^{\alpha}~.\label{eqnL41}%
\end{equation}

To shorten the notation, we henceforth eliminate the subscript \textit{diff}.
\ Because of the antisymmetry of this expression, this combination is
equivalent to one with covariant derivatives%

\begin{equation}
\xi_{\left[  1,2\right]  }^{\alpha}=\xi_{2}^{\beta}\xi_{1;\beta}^{\alpha}%
-\xi_{1}^{\beta}\xi_{2;\beta}^{\alpha}~; \label{eqnL42}%
\end{equation}
this form explicitly shows that the new parameter, $\xi_{\left[  1,2\right]
}^{\alpha}$, preserves its vector form.

Because there are no fields in parameter redefinition (\ref{eqnL41}), it
remains unaltered when we consider a double commutator, i.e.%

\begin{equation}
\xi_{\left[  \left[  1,2\right]  ,3\right]  }^{\alpha}=\xi_{3}^{\beta}%
\xi_{\left[  1,2\right]  ,\beta}^{\alpha}-\xi_{\left[  1,2\right]  }^{\beta
}\xi_{3,\beta}^{\alpha}~.\label{eqnL47}%
\end{equation}
In general, the field-independence of a new parameter is a sufficient
condition to have a group, but not a necessary condition. Should fields
appear, additional calculations of the double commutators must be performed to
find a definite answer. From (\ref{eqnL42}) one may conclude that the
\textit{diff} transformations form a group. Therefore the Jacobi identity
follows for any transformation with group properties; that is%

\begin{equation}
\left(  \left[  \left[  \delta_{2},\delta_{1}\right]  ,\delta_{3}\right]
+\left[  \left[  \delta_{3},\delta_{2}\right]  ,\delta_{1}\right]  +\left[
\left[  \delta_{1},\delta_{3}\right]  ,\delta_{2}\right]  \right)  g^{\mu\nu
}\equiv0, \label{eqnL45}%
\end{equation}
which is equivalent to a simple relationship for the gauge parameters of the
double commutators,%

\begin{equation}
\xi_{\left[  \left[  1,2\right]  ,3\right]  }^{\alpha}+\xi_{\left[  \left[
3,1\right]  ,2\right]  }^{\alpha}+\xi_{\left[  \left[  2,3\right]  ,1\right]
}^{\alpha}\equiv0. \label{eqnL46}%
\end{equation}

Note that all of the above expressions (ELDs, DIs, transformations, and even
the redefinition of parameters that support group properties) are written in
manifestly covariant form; the expectation that the results must be covariant
in a covariant theory can be used to obtain some solutions to avoid direct
calculation (e.g. construction of a covariant DI). We can use these properties
to find symmetries by using the Lagrangian approach; but a different method
can be found in the literature that has the name \textquotedblleft the
Lagrangian approach\textquotedblright\ (see \cite{Samanta} and references
therein). It proceeds by singling out one coordinate, time, followed by a long
sequence of calculations to find symmetries. For the covariant theories
discussed in \cite{Samanta} this approach has unnecessary over-complications;
the Noether DI is used as an input in this method anyway. If the Noether DI is
known, then to find a transformation requires a one-line calculation (as
(\ref{eqnL17}) - (\ref{eqnL20})), not the pages of calculation as outlined in
\cite{Samanta}. This particular \textquotedblleft Lagrangian
approach\textquotedblright\ resembles the Hamiltonian one, and it may be of
use to those who want to trace down explicit connections between the
Lagrangian and Hamiltonian methods at different stages in the calculation.

We note that Lagrangian methods are general and unrestricted by the covariance
of an action; therefore, all of the results above can be obtained without
making reference to or taking guidance from covariance, although considerable
difficulties may arise. But for any Lagrangian with an \textit{a priori}
unknown gauge symmetry, one should be able to find a DI by using only ELDs of
a given action.

According to Noether's second theorem \cite{Noether}, there is a maximum
number of independent DIs, but apparently there are no additional
restrictions. Are the DIs (\ref{eqnL15}) and (\ref{eqnL31}) for the EH action
and $\delta_{\mathit{diff}}$ transformation (\ref{eqnL33}) unique? Keeping
covariance, there is little freedom to construct a new DI (see (\ref{eqnL33}))
and its corresponding transformation; and for the EH action the covariant DI
(\ref{eqnL15}) is unique. But for the EC action there is greater freedom to
construct covariant DIs (see \cite{Trans}). If the restriction of covariance
is lifted, then the number of new combinations of DIs and new gauge
transformations become unlimited; they can be found by using a very simple
manipulation, without the need for further calculation (of course this is true
if only the transformations are of interest; but it could be a complicated
task to find, for example, a commutator like (\ref{eqnL40}), or to calculate
the Jacobi identity). If the DIs are known (e.g. (\ref{eqnL31})), one can
start to build combinations of them that are obviously also DIs. And by using
the approach of \cite{Schwinger} one may obtain the corresponding
transformations. Despite considerations of simplicity and the manifest
covariance of the DIs and transformations, are all such transformations
equally good? According to Noether's theorem, which is a general result, the
existence of a maximum number of independent DIs is an important
characteristic of a singular Lagrangian. From the rank of the Hessian of the
EH action, we know that the maximum number of independent DIs is four. So one
can obtain four new DIs by using,%

\begin{equation}
\tilde{I}_{\left(  \nu\right)  }=F_{\left(  \nu\right)  }^{\mu}\left(
g_{\alpha\beta}\right)  I_{\mu}~, \label{eqnL50}%
\end{equation}
where $I_{\mu}$ is a known DI, $\left(  \nu\right)  $ is not a covariant
index, just a numbering of the DI, and $F_{\left(  \nu\right)  }^{\mu}\left(
g_{\alpha\beta}\right)  $ are some functionals of the metric that also need
not be covariant. The only restriction on (\ref{eqnL50}) is that the
combinations be linearly independent, that is,%

\begin{equation}
\det\left\vert \frac{\partial\tilde{I}_{\left(  \nu\right)  }}{\partial
I_{\mu}}\right\vert \neq0. \label{eqnL52}%
\end{equation}

Using the approach of \cite{Schwinger}, one must consider combinations of
these four new DIs, $\tilde{I}_{\left(  \nu\right)  }$, with four gauge
functions, $\varepsilon^{\left(  \nu\right)  }$; after performing an
integration by parts, as in (\ref{eqnL17}) - (\ref{eqnL20}), one can easily
read-off the new transformations with four gauge parameters. One may equally
well perform the following rearrangements:%

\begin{equation}
\varepsilon^{\left(  \nu\right)  }\tilde{I}_{\left(  \nu\right)  }%
=\varepsilon^{\left(  \nu\right)  }F_{\left(  \nu\right)  }^{\mu}\left(
g_{\alpha\beta}\right)  I_{\mu}\equiv\tilde{\xi}^{\mu}I_{\mu}~,\label{eqnL53}%
\end{equation}
after which, transformations of the metric would have the same form as before,
but with a different, field-dependent, gauge parameter:%

\begin{equation}
\tilde{\xi}^{\mu}=\varepsilon^{\left(  \nu\right)  }F_{\left(  \nu\right)
}^{\mu}\left(  g_{\alpha\beta}\right)  .\label{eqnL54}%
\end{equation}
Therefore (\ref{eqnL54}) is a different transformation from (\ref{eqnL33});
for example, even its form cannot be preserved in calculations of the
commutators (\ref{eqnL40}) of two such transformations. The independence of
gauge parameters, stated in \cite{Hilbert, Noether, Rosenfeld}, and
\cite{Castellani}, is not contradicted by (\ref{eqnL54}), because it is merely
a short form of presentation of the results. In a full expression, the
field-independent parameter would appear ($\varepsilon^{\left(  \nu\right)  }%
$). But the \textquotedblleft quasi-covariant\textquotedblright\ form
(\ref{eqnL54}) could be useful in performing calculations. This idea will be
explained in the next Section, where one particular case of (\ref{eqnL50}%
)-(\ref{eqnL54}) is discussed: the ADM transformations.

We note that Noether's theorem and the explicit connection between DIs and
gauge transformations (as in \cite{Schwinger}) considerably simplifies the
analysis of singular Lagrangians. For example, the construction of new
transformations, as outlined above, can also be used to check the validity of
some proposed or \textquotedblleft guessed\textquotedblright\ transformations.
Assuming that a transformation is correct, one would follow \cite{Schwinger}
to construct a corresponding DI candidate that can be checked by direct
substitution of the ELDs. If the candidate is a true DI, then by the converse
of Noether's theorem, it is a symmetry. By checking an identity one may manage
expressions of any complexity because terms of different types can be
considered separately; this property is very important for dealing with
non-covariant expressions (i.e. all terms with a particular derivative of a
particular field should be zero independently of the rest of an expression).
This method is simpler if some DIs are already known; in such a case, one may
express the new DIs as combinations of known identities, as in (\ref{eqnL53})
(e.g. see \cite{BGMR}), which is sufficient confirmation of the correctness of
the proposed transformations.

\section{ADM symmetry of the EH action}

The transformations of the ADM variables, $\delta_{\mathit{ADM}}\left\{
N,N^{i},\gamma_{km}\right\}  $, that follow from the constraints of the ADM
Hamiltonian are well-known; and using (\ref{eqnL1}) allows one to find the
transformations of the metric tensor, $\delta_{\mathit{ADM}}\left\{  g^{\mu
\nu}\right\}  $. They can be presented in the following form:%

\begin{equation}
\delta_{ADM}g^{\mu\nu}=\tilde{\xi}_{,\alpha}^{\nu}g^{\mu\alpha}+\tilde{\xi
}_{,\alpha}^{\mu}g^{\nu\alpha}-g_{,\alpha}^{\mu\nu}\tilde{\xi}^{\alpha}
\label{eqnL60}%
\end{equation}
with $\tilde{\xi}^{\alpha}$ given by\ %

\begin{equation}
\tilde{\xi}^{\nu}=\delta_{0}^{\nu}\left(  -g^{00}\right)  ^{\frac{1}{2}%
}\varepsilon^{\perp}+\delta_{k}^{\nu}\left[  \varepsilon^{k}+\frac{g^{0k}%
}{g^{00}}\left(  -g^{00}\right)  ^{\frac{1}{2}}\varepsilon^{\perp}\right]
\label{eqnL61}%
\end{equation}
(e.g. see appendix of \cite{Castellani}, for more detailed calculations in
\cite{Myths} and also \cite{Bergmann}).

This representation helps one to see the origin of some of the names of the
ADM transformations: \textquotedblleft specific metric-dependent
diffeomorphisms\textquotedblright\ \cite{Pons},\ or \textquotedblleft a
one-to-one correspondence between the diffeomorphisms and the gauge
variations\textquotedblright\ \cite{Mukherjee}, or \textquotedblleft
diffeomorphism-induced gauge symmetry\textquotedblright\ \cite{PonsSS}. From
the discussion at the end of the previous Section, one may see that
(\ref{eqnL61}) is one of many possible \textquotedblleft field-dependent
diffeomorphisms\textquotedblright\ and \textquotedblleft one-to-one
correspondence\textquotedblright. Using the transformations (\ref{eqnL60}),
the corresponding DIs can be restored. Because they are combinations of the
known covariant DIs and they are also the identities, the transformations
(\ref{eqnL60}) represent the gauge symmetry of the EH Lagrangian. So whatever
the field dependence of the transformations of the form shown in
(\ref{eqnL61}) might be, these transformations are guaranteed to be a symmetry
of the EH Lagrangian. We can also explicitly find separate identities that
correspond to each parameter ($\varepsilon^{\perp}$, $\varepsilon^{k}$) of the
ADM transformations:%

\begin{equation}
\tilde{\xi}^{\nu}I_{\nu}=\varepsilon^{\perp}\left[  \frac{g^{0k}}{g^{00}%
}\left(  -g^{00}\right)  ^{\frac{1}{2}}I_{k}+\left(  -g^{00}\right)
^{\frac{1}{2}}I_{0}\right]  +\varepsilon^{k}I_{k}~, \label{eqnL62}%
\end{equation}
which in turn, give two DIs to describe the ADM transformations:%

\[
\tilde{\xi}^{\nu}I_{\nu}=\varepsilon^{\perp}\tilde{I}_{\bot}+\varepsilon
^{k}\tilde{I}_{k}~,
\]
with%

\begin{equation}
\tilde{I}_{\bot}=\frac{g^{0k}}{g^{00}}\left(  -g^{00}\right)  ^{\frac{1}{2}%
}I_{k}+\left(  -g^{00}\right)  ^{\frac{1}{2}}I_{0}~, \label{eqnL63}%
\end{equation}
and%
\begin{equation}
\tilde{I}_{k}=I_{k}~. \label{eqnL64}%
\end{equation}
These are obviously DIs since they are linear combinations of the components
of the covariant DI.

The names \textquotedblleft field-dependent diffeomorphism\textquotedblright%
\ and \textquotedblleft one-to-one correspondence\textquotedblright\ are
misleading. This transformation is different from diffeomorphism and even its
resemblance to \textit{diff} in \textquotedblleft form\textquotedblright%
\ disappears if one were to calculate the commutator of two such
transformations. The previous relation for \textit{diff} (\ref{eqnL41})
changes and a simple substitution of (\ref{eqnL62}) into (\ref{eqnL41}) is not
equivalent to the direct calculation of the commutator%

\[
\tilde{\xi}_{\left[  1,2\right]  }^{\alpha}\neq\tilde{\xi}_{2}^{\beta}%
\tilde{\xi}_{1,\beta}^{\alpha}-\tilde{\xi}_{2}^{\beta}\tilde{\xi}_{1,\beta
}^{\alpha}%
\]
in which extra contributions appear. This was noticed in \cite{PonsSS}:
\textquotedblleft It is impossible to get for $\xi_{3}$ [our $\tilde{\xi
}_{\left[  1,2\right]  }^{\alpha}$] the standard diffeomorphism
rule\textquotedblright\ and so the transformation with parameters
(\ref{eqnL61}) is not a field-dependent diffeomorphism, but a different
symmetry. Even the formal resemblance of \textit{diff} transformations does
not survive in the commutator.

Let us try to find the commutator of two ADM transformations. From now on we
shall eliminate the subscript ADM, and use $\delta_{ADM}g^{\mu\nu}%
=\tilde{\delta}g^{\mu\nu}$ to abbreviate the notation. The quasi-covariant
form of (\ref{eqnL61}) allows one to simplify the calculations by using some
of the results from the previous Section, and then to consider the
transformations of all of the components of a contravariant tensor at once;
this is impossible to do when the ADM Lagrangian is analyzed.

In performing the calculation of the commutator of the ADM transformation,
that is%

\[
\left(  \tilde{\delta}_{2}\tilde{\delta}_{1}-\tilde{\delta}_{1}\tilde{\delta
}_{2}\right)  g^{\mu\nu}=\tilde{\delta}_{2}\left(  \tilde{\xi}_{1,\alpha}%
^{\nu}g^{\mu\alpha}+\tilde{\xi}_{1,\alpha}^{\mu}g^{\nu\alpha}-g_{,\alpha}%
^{\mu\nu}\tilde{\xi}_{1}^{\alpha}\right)  -\tilde{\delta}_{1}\left(
\tilde{\xi}_{2,\alpha}^{\nu}g^{\mu\alpha}+\tilde{\xi}_{2,\alpha}^{\mu}%
g^{\nu\alpha}-g_{,\alpha}^{\mu\nu}\tilde{\xi}_{2}^{\alpha}\right)  ,
\]
the result found differs from that obtained by calculating the commutator of
the diffeomorphism transformations; this difference exists because of the
presence of the fields in $\tilde{\xi}^{\alpha}$ (which is an abbreviated form
(\ref{eqnL61}), not a field-independent parameter). Consider%

\[
\left(  \tilde{\delta}_{2}\tilde{\delta}_{1}-\tilde{\delta}_{1}\tilde{\delta
}_{2}\right)  g^{\mu\nu}=\tilde{\xi}_{1,\alpha}^{\nu}\delta_{2}g^{\mu\alpha
}+\tilde{\xi}_{1,\alpha}^{\mu}\delta_{2}g^{\nu\alpha}-\left(  \delta_{2}%
g^{\mu\nu}\right)  _{,\alpha}\tilde{\xi}_{1}^{\alpha}-\tilde{\xi}_{2,\alpha
}^{\nu}\delta_{1}g^{\mu\alpha}-\tilde{\xi}_{2,\alpha}^{\mu}\delta_{1}%
g^{\nu\alpha}+\left(  \delta_{1}g^{\mu\nu}\right)  _{,\alpha}\tilde{\xi}%
_{2}^{\alpha}%
\]

\[
+\left(  \delta_{2}\tilde{\xi}_{1}^{\nu}\right)  _{,\alpha}g^{\mu\alpha
}+\left(  \delta_{2}\tilde{\xi}_{1}^{\mu}\right)  _{,\alpha}g^{\nu\alpha
}-g_{,\alpha}^{\mu\nu}\delta_{2}\tilde{\xi}_{1}^{\alpha}-\left(  \delta
_{1}\tilde{\xi}_{2}^{\nu}\right)  _{,\alpha}g^{\mu\alpha}-\left(  \delta
_{1}\tilde{\xi}_{2}^{\mu}\right)  _{,\alpha}g^{\nu\alpha}+g_{,\alpha}^{\mu\nu
}\delta_{1}\tilde{\xi}_{2}^{\alpha}~;
\]
the terms in the first line (no contributions with $\tilde{\delta}\tilde{\xi
}^{\alpha}$) give the same result as that for the known \textit{diff} (with
$\tilde{\xi}^{\alpha}$); the second line produces additional contributions
that can be combined into the following form:%

\[
\left(  \delta_{2}\tilde{\xi}_{1}^{\nu}-\delta_{1}\tilde{\xi}_{2}^{\nu
}\right)  _{,\alpha}g^{\mu\alpha}+\left(  \delta_{2}\tilde{\xi}_{1}^{\mu
}-\delta_{1}\tilde{\xi}_{2}^{\mu}\right)  _{,\alpha}g^{\nu\alpha}-g_{,\alpha
}^{\mu\nu}\left(  \delta_{2}\tilde{\xi}_{1}^{\alpha}-\delta_{1}\tilde{\xi}%
_{2}^{\alpha}\right)  .
\]
After making some simple rearrangement, we obtain a general expression:%

\begin{equation}
\tilde{\xi}_{\left[  1,2\right]  }^{\alpha}=\tilde{\xi}_{2}^{\beta}\tilde{\xi
}_{1,\beta}^{\alpha}-\tilde{\xi}_{2}^{\beta}\tilde{\xi}_{1,\beta}^{\alpha
}+\delta_{2}\tilde{\xi}_{1}^{\alpha}-\delta_{1}\tilde{\xi}_{2}^{\alpha}
\label{eqnL72}%
\end{equation}
with additional contributions that must be explicitly calculated for the
particular field dependence of the parameters.

In the first two terms of (\ref{eqnL72}), we merely substitute (\ref{eqnL61});
and in the last two terms (which are zero if parameters are \textquotedblleft
field-independent\textquotedblright),\ we have%

\begin{equation}
\delta_{2}\tilde{\xi}_{1}^{\alpha}-\delta_{1}\tilde{\xi}_{2}^{\alpha}%
=\delta_{0}^{\alpha}\varepsilon_{1}^{\perp}\delta_{2}\left(  -g^{00}\right)
^{\frac{1}{2}}+\delta_{k}^{\alpha}\varepsilon_{1}^{\perp}\delta_{2}\left[
\frac{g^{0k}}{g^{00}}\left(  -g^{00}\right)  ^{\frac{1}{2}}\right]
-\delta_{0}^{\alpha}\varepsilon_{2}^{\perp}\delta_{1}\left(  -g^{00}\right)
^{\frac{1}{2}}-\delta_{k}^{\alpha}\varepsilon_{2}^{\perp}\delta_{1}\left[
\frac{g^{0k}}{g^{00}}\left(  -g^{00}\right)  ^{\frac{1}{2}}\right]  .
\label{eqnL73}%
\end{equation}
(Note that in (\ref{eqnL73}) $\varepsilon^{k}$ is absent, since it enters
(\ref{eqnL61}) without field-dependent coefficients.) The final result for
(\ref{eqnL72}) can be presented in the same form as (\ref{eqnL60}),%

\[
\tilde{\xi}_{\left[  1,2\right]  }^{\alpha}=\delta_{0}^{\alpha}\left(
-g^{00}\right)  ^{\frac{1}{2}}\varepsilon_{\left[  1,2\right]  }^{\perp
}+\delta_{k}^{\alpha}\left[  \varepsilon_{\left[  1,2\right]  }^{k}%
+\frac{g^{0k}}{g^{00}}\left(  -g^{00}\right)  ^{\frac{1}{2}}\varepsilon
_{\left[  1,2\right]  }^{\perp}\right]
\]
where%

\begin{equation}
\varepsilon_{\left[  1,2\right]  }^{\perp}=\varepsilon_{2}^{k}\varepsilon
_{1,k}^{\bot}-\varepsilon_{1}^{k}\varepsilon_{2,k}^{\bot} \label{eqnL80}%
\end{equation}
and%
\begin{equation}
\varepsilon_{\left[  1,2\right]  }^{k}=\varepsilon_{2}^{m}\varepsilon
_{1,m}^{k}-\varepsilon_{1}^{m}\varepsilon_{2,m}^{k}+\left(  \varepsilon
_{1,m}^{\bot}\varepsilon_{2}^{\bot}-\varepsilon_{2,m}^{\bot}\varepsilon
_{1}^{\bot}\right)  e^{mk}. \label{eqnL81}%
\end{equation}

Here the combination, $e^{mk}$, which found in Dirac's Hamiltonian analysis of
the EH action, is formed%

\[
e^{mk}=g^{mk}-\frac{g^{0m}g^{0k}}{g^{00}}\text{ \ \ \ }g_{nm}e^{mk}=\delta
_{n}^{k}.
\]
Due to the presence of fields in (\ref{eqnL81}), one might conclude that this
\textquotedblleft soft\ algebra\textquotedblright\ structure signifies that
the symmetry transformations no longer form a group. This is a possible
outcome when fields appear in the structure constant, but not always. The
field independence of the parameters in a commutator of two transformations is
a sufficient condition to have an algebra, but not a necessary one.

With the appearance of fields, such as in (\ref{eqnL81}), the double
commutator must be checked by direct calculation. Again, we can use the
general form of the results and consider the double commutator. We return to
(\ref{eqnL72}), which is a general expression whatever the field dependence of
the gauge parameters might be, and by making the changes $1\rightarrow\left[
1,2\right]  $ and $2\rightarrow3$, we obtain%

\begin{equation}
\tilde{\xi}_{\left[  \left[  1,2\right]  ,3\right]  }^{\alpha}=\tilde{\xi}%
_{3}^{\beta}\tilde{\xi}_{\left[  1,2\right]  ,\beta}^{\alpha}-\tilde{\xi
}_{\left[  1,2\right]  }^{\beta}\tilde{\xi}_{3,\beta}^{\alpha}+\delta
_{3}\tilde{\xi}_{\left[  1,2\right]  }^{\alpha}-\delta_{\left[  1,2\right]
}\tilde{\xi}_{3}^{\alpha}. \label{eqnL85}%
\end{equation}
The evaluation of the first two terms is straightforward; but the second pair,
because of the presence of $\xi_{\left[  1,2\right]  }^{\alpha}$ (with fields,
see (\ref{eqnL81})), produces an additional contribution as compared to the
simple change of indices ($1\rightarrow\left[  1,2\right]  $ and
$2\rightarrow3$) in (\ref{eqnL73}):%

\[
\delta_{3}\tilde{\xi}_{\left[  1,2\right]  }^{\alpha}-\delta_{\left[
1,2\right]  }\tilde{\xi}_{3}^{\alpha}=\delta_{0}^{\alpha}\varepsilon_{\left[
1,2\right]  }^{\perp}\delta_{3}\left(  -g^{00}\right)  ^{\frac{1}{2}}%
+\delta_{k}^{\alpha}\varepsilon_{\left[  1,2\right]  }^{\perp}\delta
_{3}\left[  \frac{g^{0k}}{g^{00}}\left(  -g^{00}\right)  ^{\frac{1}{2}%
}\right]  +\delta_{k}^{\alpha}\delta_{3}\varepsilon_{\left[  1,2\right]  }^{k}%
\]

\begin{equation}
-\delta_{0}^{\alpha}\varepsilon_{3}^{\perp}\delta_{\left[  1,2\right]
3}\left(  -g^{00}\right)  ^{\frac{1}{2}}-\delta_{k}^{\alpha}\varepsilon
_{3}^{\perp}\delta_{\left[  1,2\right]  }\left[  \frac{g^{0k}}{g^{00}}\left(
-g^{00}\right)  ^{\frac{1}{2}}\right]  . \label{eqnL86}%
\end{equation}

The last contribution in the first line was absent from (\ref{eqnL73}) because
$\varepsilon^{k}$ is all field-independent. This additional contribution is%

\[
\hat{\varepsilon}_{\left[  \left[  1,2\right]  ,3\right]  }^{k}=\delta
_{3}\varepsilon_{\left[  1,2\right]  }^{k}=\left(  \varepsilon_{1,m}^{\bot
}\varepsilon_{2}^{\bot}-\varepsilon_{2,m}^{\bot}\varepsilon_{1}^{\bot}\right)
\delta_{3}e^{mk}.
\]
After performing a transformation $\delta_{3}e^{mk}$, it leads to%

\begin{equation}
\hat{\varepsilon}_{\left[  \left[  1,2\right]  ,3\right]  }^{k}=\left(
\varepsilon_{1,m}^{\bot}\varepsilon_{2}^{\bot}-\varepsilon_{2,m}^{\bot
}\varepsilon_{1}^{\bot}\right)  \left\{  \varepsilon_{3,n}^{m}e^{kn}%
+\varepsilon_{3,n}^{k}e^{mn}-e_{,n}^{km}\varepsilon_{3}^{n}\right.
\label{eqnL87}%
\end{equation}

\[
\left.  +\left(  -g^{00}\right)  ^{\frac{1}{2}}\left[  \left(  \frac{g^{0k}%
}{g^{00}}\right)  _{,n}e^{mn}+\left(  \frac{g^{0m}}{g^{00}}\right)
_{,n}e^{kn}-e_{,n}^{km}\frac{g^{0n}}{g^{00}}-e_{,0}^{km}\right]
\varepsilon_{3}^{\perp}\right\}  .
\]
The remaining contributions (see  (\ref{eqnL85}) and (\ref{eqnL86})) are the
same as those found in the previous calculations, so we can use (\ref{eqnL81})
with ($1\rightarrow\left[  1,2\right]  $ and $2\rightarrow3$), as before, to obtain:%

\begin{equation}
\varepsilon_{\left[  \left[  1,2\right]  ,3\right]  }^{k}=-\varepsilon
_{\left[  1,2\right]  }^{i}\varepsilon_{3,i}^{k}+\varepsilon_{3}%
^{i}\varepsilon_{\left[  1,2\right]  ,i}^{k}+\varepsilon_{\left[  1,2\right]
,m}^{\bot}\varepsilon_{3}^{\bot}e^{mk}-\varepsilon_{3,m}^{\bot}\varepsilon
_{\left[  1,2\right]  }^{\bot}e^{mk}. \label{eqnL88}%
\end{equation}
After the substitution of $\varepsilon_{\left[  1,2\right]  }^{i}$, we have:%

\[
\varepsilon_{\left[  \left[  1,2\right]  ,3\right]  }^{k}=-\left(
-\varepsilon_{1}^{m}\varepsilon_{2,m}^{i}+\varepsilon_{2}^{m}\varepsilon
_{1,m}^{i}+\varepsilon_{1,m}^{\bot}\varepsilon_{2}^{\bot}e^{mi}-\varepsilon
_{2,m}^{\bot}\varepsilon_{1}^{\bot}e^{mi}\right)  \varepsilon_{3,i}^{k}%
\]

\[
+\varepsilon_{3}^{i}\left(  -\varepsilon_{1}^{m}\varepsilon_{2,m}%
^{k}+\varepsilon_{2}^{m}\varepsilon_{1,m}^{k}+\varepsilon_{1,m}^{\bot
}\varepsilon_{2}^{\bot}e^{mk}-\varepsilon_{2,m}^{\bot}\varepsilon_{1}^{\bot
}e^{mk}\right)  _{,i}%
\]

\begin{equation}
+\left(  -\varepsilon_{1}^{n}\varepsilon_{2,n}^{\bot}+\varepsilon_{2}%
^{n}\varepsilon_{1,n}^{\bot}\right)  _{,m}\varepsilon_{3}^{\bot}%
e^{mk}-\varepsilon_{3,m}^{\bot}\left(  -\varepsilon_{1}^{n}\varepsilon
_{2,n}^{\bot}+\varepsilon_{2}^{n}\varepsilon_{1,n}^{\bot}\right)  e^{mk}.
\label{eqnL89}%
\end{equation}

Combining (\ref{eqnL87}) and (\ref{eqnL89}) leads to some simplification; but
the condition, which must be satisfied for the Jacobi identities to be
correct, does not hold:%

\[
\varepsilon_{\left[  \left[  1,2\right]  ,3\right]  }^{k}+\varepsilon_{\left[
\left[  2,3\right]  ,1\right]  }^{k}+\varepsilon_{\left[  \left[  3,1\right]
,2\right]  }^{k}\neq0
\]
(one contribution that prevents cancellation is the term in (\ref{eqnL87})
proportional to $e_{,n}^{km}\varepsilon_{3}^{n}$).

The EH action is invariant under the ADM transformations, but unlike
\textit{diff}, $\delta_{\mathit{ADM}}\left\{  g^{\mu\nu}\right\}  $ do not
form a group. \ This result illustrates that all possible symmetries, which
can be constructed easily from various combinations of DIs, are not equally
good. There is one transformation (in  general, some restricted class of
transformations) that forms a group; and such transformations constitute the
\textquotedblleft basic\textquotedblright\ or \textquotedblleft
true\textquotedblright\ gauge symmetry of the Lagrangian. \ In analogy with
the Hamiltonian formulation, one might call a symmetry that can form a group a
\textquotedblleft canonical\textquotedblright\ symmetry of the Lagrangian.

\section{Conclusion}

The application of Dirac's method to derive the Hamiltonian formulations of
the EH Lagrangian, $L_{EH}\left(  g^{\mu\nu}\right)  $, and the ADM
Lagrangian, $L_{ADM}\left(  N,N^{i},\gamma_{km}\right)  $, leads to two
different gauge symmetries; because of this difference in symmetries, it is no
surprise that their Hamiltonian formulations are not related by a canonical
transformation \cite{FKK, Myths}. If the Hamiltonian method is considered to
be an algorithm that allows one to restore a gauge symmetry, and if the
Lagrangian and Hamiltonian methods are equivalent, then one might conclude
that the two Lagrangians are not equivalent \cite{Myths}. The expression
"non-canonicity puzzle" was coined to describe this result\ \cite{CLM}. But if
equivalence of two Lagrangians is assumed, then one might alternatively
conclude that the Hamiltonian method is not an algorithm (at least in its
currently known form or for this particular case); thus Dirac's method must be
modified \cite{ShestakovaGandC, ShestakovaCQG}.

In this paper we offer a preliminary answer to the question of how to compare
the symmetries of two Lagrangians which differ by invertible change of
variables. Before such an undertaking is made, it is essential to understand
how to distinguish two symmetries for the same Lagrangian. Based on Noether's
theorem, we demonstrate that both symmetries (\textit{diff} and ADM) are
symmetries of the EH Lagrangian, when written for the same variables; we also
demonstrate that more symmetries can be constructed using the Lagrangian
method. But a study of their group properties reveals that only one symmetry,
\textit{diff,}\ has group properties; and neither ADM nor any other
symmetries, constructed by using a so-called field-dependent redefinition of
gauge parameters, have such a property. Therefore, for the EH Lagrangian, only
one distinct symmetry with a group property exists (canonical symmetry).

To call two Lagrangians equivalent, any and all canonical symmetries should be
presented in both formulations. The ADM symmetry, which follows from the
Hamiltonian formulation of the ADM action, is not a canonical symmetry of the
EH action. Of course, the question whether the ADM formulation possesses
canonical symmetry needs to be answered. Such calculations are
straightforward, but extremely cumbersome (the penalty for working with
non-covariant variables); and the relatively simple calculations presented in
this article, which use a quasi-covariant form to allow one to consider
transformations for all components of metric at once, are impossible in the
case of the ADM Lagrangian. The calculations must be performed separately for
all fields, and the redefinition of the gauge parameters must be the same for
all fields. The DIs are also much more complicated, especially for the
transformation of the ADM variables under diffeomorphism\footnote{In addition,
such DIs are not covariant, and because of this, cannot be true in all
coordinate systems.}; and such transformations must also be checked to
determine if they correspond to symmetries with a group property for the ADM Lagrangian.

From the analysis of the invariance of the EH Lagrangian performed in this
paper, it follows that $\delta_{\mathit{diff}}$ has a group property; but
$\delta_{\mathit{ADM}}$ does not. We are currently undertaking an
investigation of the properties of these two symmetries for the ADM
Lagrangian. There are four possible cases, all of which lead to contradictions
and further questions. For the ADM Lagrangian, these cases are:

(a) both transformations form groups;

(b) neither transformation forms a group;

(c) $\delta_{\mathit{ADM}}$ forms a group, but not $\delta_{\mathit{diff}}$;

(d) $\delta_{\mathit{diff}}$ forms a group, but not $\delta_{\mathit{ADM}}$.

The first three of these cases lead to the non-equivalence of the Lagrangians.
Cases (a) and (b) raise a question about the uniqueness of Dirac's procedure.
The two transformations both form groups (case (a)), or neither of them forms
a group (case (b)); but only one symmetry is chosen by the Hamiltonian
procedure. Case (c) is consistent with the uniqueness of the Hamiltonian
method, as for the EH action, it selects a symmetry with a group property; but
the Lagrangians (ADM and EH) cannot be equivalent.

Case (d) would imply an equivalence of the canonical symmetries of the ADM and
EH Lagrangians, and that \textit{diff} is a symmetry with group properties for
the ADM Lagrangian; but such a conclusion contradicts the widely quoted
statement of Isham and Kuchar \cite{Isham}: \textquotedblleft the full group
of spacetime diffeomorphisms has \textit{somehow} got lost in making the
transition from the Hilbert action to the Dirac-ADM action\textquotedblright%
\ (italic is ours)\footnote{The name of Dirac is used incorrectly in this
statement because Dirac's Hamiltonian is not canonically related to the ADM
Hamiltonian and, in addition, Dirac's modification of the EH action is
performed in the way to preserve Einstein's equations. Moreover, if case (d)
is correct, then neither for the Dirac nor for the ADM action the
diffeomorphism \textquotedblleft got lost\textquotedblright.}.\ Such
statements were based on the results of the Hamiltonian formulation of the ADM
Lagrangian with ADM gauge transformations. And one can often find claims that
only spatial \textit{diff} is a symmetry of the ADM formulation. (Such
statements are not compatible at all with equivalence of the EH and ADM
actions.) So, case (d) would inexorably lead one to conclude that Dirac's
Hamiltonian method does not work for ADM variables (for the metric formulation
of the EH action, it picks the symmetry with group properties, but for the ADM
action it fails to do so). This outcome would force one to reconsider the
\textquotedblleft theoretical foundations\textquotedblright;\ to be more
precise, to reconsider Dirac's method, as suggested in \cite{ShestakovaGandC,
ShestakovaCQG}, and to doubt its validity as an algorithm (at least in its
current form). An algorithm should work without an \textit{a priori} knowledge
of the gauge symmetry, and not demand modification of the method to adjust its
outcome to the results that are known, \textit{a priori,} for a particular
Lagrangian (e.g. ADM). Note also that such a formulation should be expected to
be connected by a canonical transformation to the Hamiltonians of PSS and
Dirac. We plan to continue this discussion after completing the analysis of
the group properties of two transformations for the ADM Lagrangian.

There is another solution to the \textquotedblleft puzzle\textquotedblright%
,\ but it would probably not be well accepted or considered seriously in view
of the movement to devalue the importance of general covariance. This
historical change of views on covariance is expressed perfectly by Norton
\cite{Norton}: \textquotedblleft When Einstein formulated his General Theory
of Relativity, he presented it as the culmination of his search for a
generally covariant theory. That this was the signal achievement of the theory
rapidly became the orthodox conception. A dissident view, however, tracing
back at least to objections raised by Eric Kretschmann in 1917, holds that
there is no physical content in Einstein's demand for general covariance. That
dissident view has grown into the mainstream. Many accounts of general
relativity no longer even mention a principle or requirement of general
covariance.\textquotedblright\ 

Considering the EH action and its original variables (the metric tensor), the
Hamiltonian method (innately non-covariant) or combinations of DIs (which can
be chosen to be unrestricted by covariance) both single out the one unique,
covariant symmetry. Covariance is neither demanded nor encoded in either of
these methods; but when they are applied to covariant actions only covariant
results are \textit{\textquotedblleft somehow\textquotedblright} produced.
Many statements can be found in the literature that are similar to the recent
one in \cite{HKG2010}: \textquotedblleft one of the beauties of general
relativity is that it is difficult to deform it without running into
inconsistencies\textquotedblright. Maybe, the solution to the
\textquotedblleft puzzle\textquotedblright\ is simple: do not destroy
covariance - \textquotedblleft one of the beauties\textquotedblright\ of
Einstein's theory; and do not deform it by using non-covariant variables.
Heeding these caveats will prevent one from \textquotedblleft running into
inconsistencies\textquotedblright, finding contradictions, and facing such
\textquotedblleft puzzles\textquotedblright. Further, instead of being on the
horns of a dilemma, to choose \textquotedblleft canonical or
covariant\textquotedblright\ \cite{CLM},\ one might simply conclude: only
covariant results are canonical for General Relativity.

\section{Acknowledgment}

We would like to thank A. Frolov, L.A. Komorowski, D.G.C. McKeon, and A.V.
Zvelindovsky for discussions.

\end{document}